\documentclass[a4paper,english,conference]{IEEEtran}
\usepackage{float}
\usepackage{latexsym}

\usepackage{amsfonts}
\usepackage{amsbsy}
\usepackage{amssymb}
\usepackage{times}
\usepackage{graphicx}
\usepackage{setspace}
\usepackage{enumerate}
\usepackage[usenames]{color}
\usepackage[dvips]{pstcol}
\usepackage{epstopdf}
\usepackage[caption=false]{subfig}
\usepackage{cite}
\usepackage{amssymb}
\usepackage{amsfonts}
\usepackage{graphicx}
\usepackage{epsfig}
\usepackage{psfrag}
\usepackage{xcolor}
\usepackage{amsfonts, bm}
\usepackage{epstopdf}
\usepackage{cite}
\usepackage{color}
\usepackage{xcolor}
\usepackage{subfig}
\usepackage{verbatim}
\usepackage{multirow}
\usepackage{booktabs}
\usepackage{amsthm}
\usepackage{makecell}
\usepackage{units}
\usepackage[linesnumbered, ruled]{algorithm2e}
\usepackage{algpseudocode}
\usepackage{amsmath}

\usepackage[linesnumbered, ruled]{algorithm2e}
\usepackage{algpseudocode}
\usepackage{amsmath}

\IEEEoverridecommandlockouts

\usepackage{geometry}
\begin{document}
	
	\title{Joint Transmission and Deblurring: A Semantic Communication Approach Using Events}
			%
	\author{\IEEEauthorblockN{Pujing Yang \IEEEauthorrefmark{1}, Guangyi Zhang \IEEEauthorrefmark{1}, Yunlong Cai \IEEEauthorrefmark{1}, Lei Yu \IEEEauthorrefmark{2}, and Guanding Yu \IEEEauthorrefmark{1}}
		\IEEEauthorblockA{\IEEEauthorrefmark{1} College of Information Science and Electronic Engineering, Zhejiang University, Hangzhou, China }
		\IEEEauthorblockA{\IEEEauthorrefmark{2} School of Electronic Information,
			Wuhan University, Wuhan, China  \\ E-mail: \{yangpujing, zhangguangyi, ylcai, yuguanding\}@zju.edu.cn, ly.wd@whu.edu.cn} }

	\maketitle
	\vspace{-3.3em}
	
	\begin{abstract}
		Deep learning-based joint source-channel coding (JSCC) is emerging as a promising technology for effective image transmission. However, most existing approaches focus on transmitting clear images, overlooking real-world challenges such as motion blur caused by camera shaking or fast-moving objects. Motion blur often degrades image quality, making transmission and reconstruction more challenging. Event cameras, which asynchronously record pixel intensity changes with extremely low latency, have shown great potential for motion deblurring tasks. However, the efficient transmission of the abundant data generated by event cameras remains a significant challenge.
		In this work, we propose a novel JSCC framework for the joint transmission of blurry images and events, aimed at achieving high-quality reconstructions under limited channel bandwidth. This approach is designed as a deblurring task-oriented JSCC system. Since RGB cameras and event cameras capture the same scene through different modalities, their outputs contain both shared and domain-specific information. To avoid repeatedly transmitting the shared information, we extract and transmit their shared information and domain-specific information, respectively. At the receiver, the received signals are processed by a deblurring decoder to generate clear images. Additionally, we introduce a multi-stage
		training strategy to train the proposed model. Simulation results demonstrate that our method significantly outperforms existing JSCC-based image transmission schemes, addressing motion blur effectively.
		
	\end{abstract}
	
	
	\IEEEpeerreviewmaketitle
	
	\section{Introduction}
	The rapid expansion of ultra-large-scale image and video transmission has intensified the challenges posed by limited bandwidth conditions \cite{Xinchao_Access2024, zhang2024MDJCM, AIenable}. To address this issue, deep learning-based joint source-channel coding (JSCC) for semantic communications has garnered significant attention \cite{Guangyi_TCOM2024, Eirina_TCCN2019, ADJSCC}. As defined in \cite{threelevel}, semantic communications represent the second level of communication, focusing on extracting and transmitting the most critical task-oriented features of source data. This approach significantly enhances the efficiency of channel bandwidth utilization, making it a promising solution for bandwidth-constrained scenarios.

	\begin{figure}[t]
		\begin{centering}
			\subfloat[]{\label{event_blurry}\includegraphics[width=6.2cm]{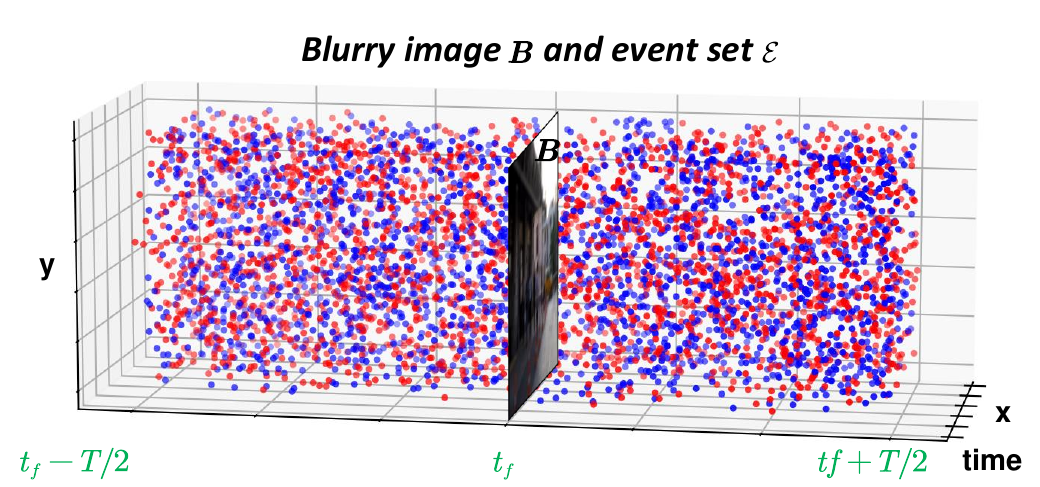}} \\
			\subfloat[]{\label{blurry}\includegraphics[width=4.2cm]{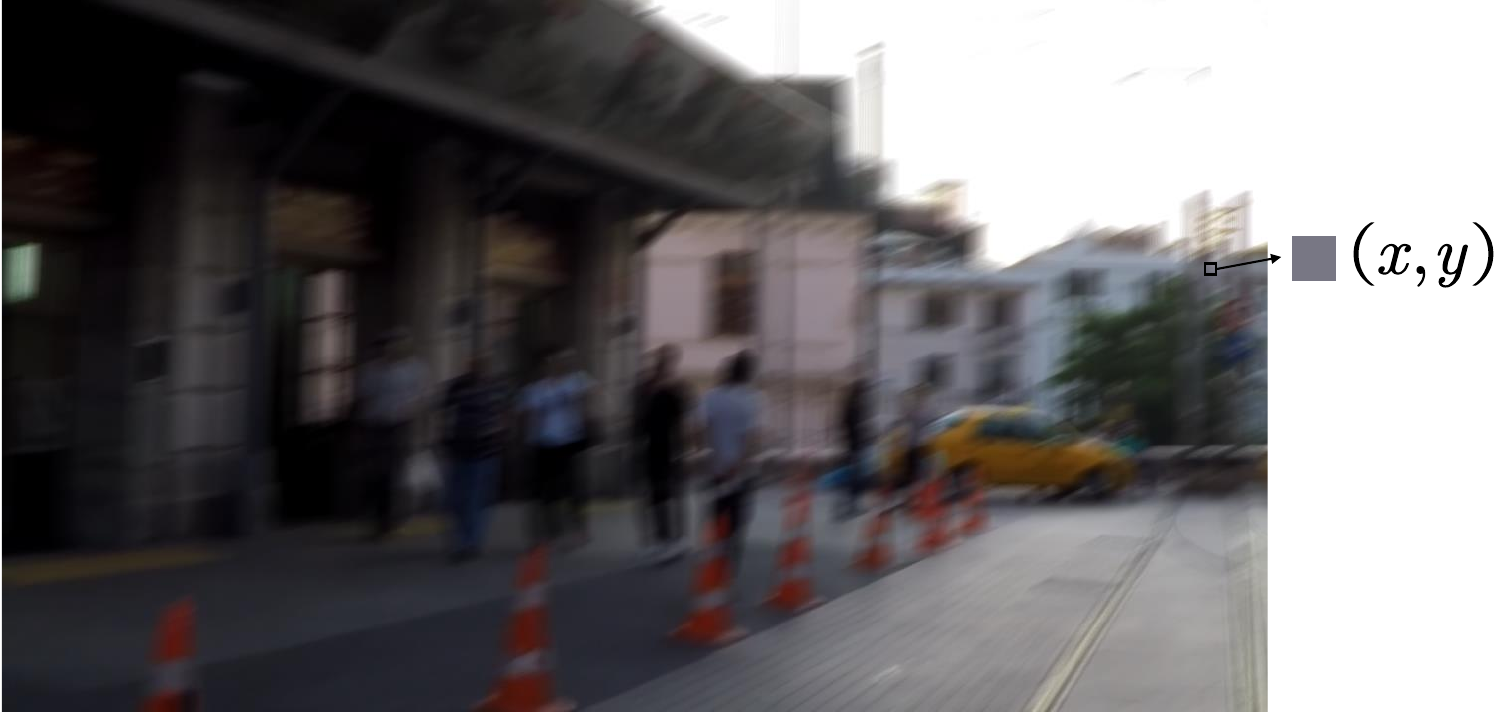}}
			\subfloat[]{\label{events}\includegraphics[width=4.2cm]{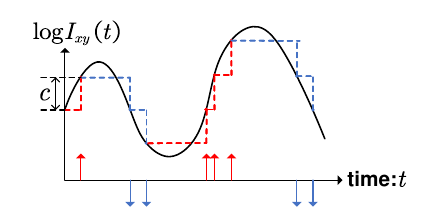}}
			\caption{(a) A blurry image $\boldsymbol{B}$ with exposure period [$t_f-T/2$, $t_f+T/2$] and events $\mathcal{E}$ triggered during this period, where red and blue dots represent positive and negative events, respectively; (b) A larger version of the blurry image $\boldsymbol{B}$; (c) The produces of events for a specific pixel ($x,y$) in (b). When the logarithm intensity change exceeds a threshold $c$, the event camera sends an event (i.e., intensity increase results in a positive event (plotted in red) while intensity decrease results in a negative event (plotted in blue)). The generated event set is $\mathcal{E}_{xy}=\{ e_i = (x,y,t_i,p_i):0 \leq i \leq 7 \}$.}
			\label{blurry_events}
		\end{centering}
	\end{figure}	
	\begin{figure*}[t]
		\begin{centering}
			\includegraphics[width=0.93 \textwidth]{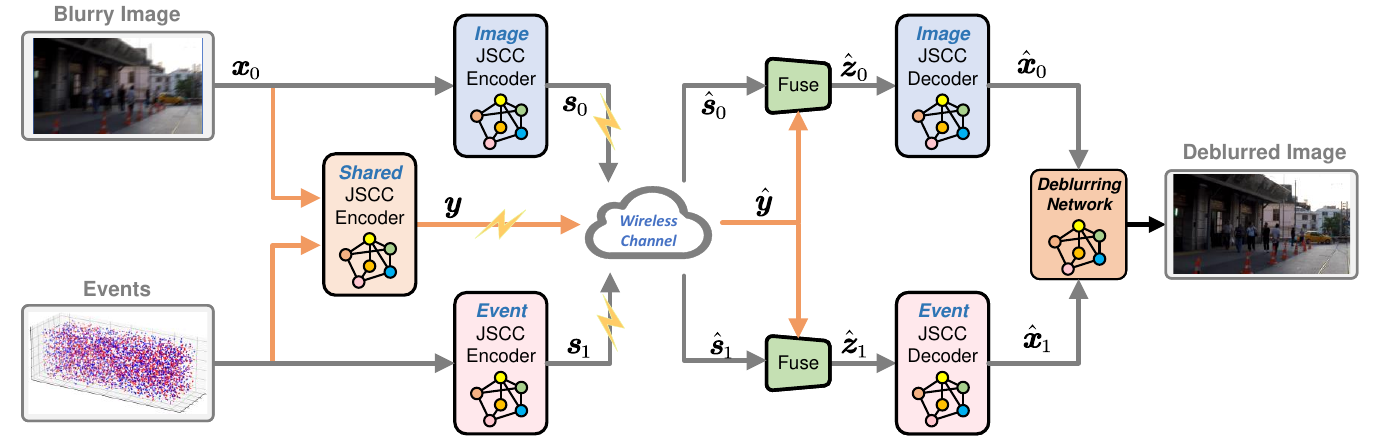}
			\par \end{centering}
		\caption{The framework of the proposed EV-JSCC. }
		\label{framework}
	\end{figure*}
	While semantic communications have been applied successfully to wireless image transmission, most existing work predominantly focuses on transmitting clear images \cite{Guangyi_TCOM2024, Eirina_TCCN2019, ADJSCC}.
	However, real-world scenarios often involve capturing images plagued by motion blur caused by factors such as camera shake or rapid object movement, as illustrated in Fig. \ref{blurry_events}(b). 
	For instance, a smartphone camera capturing scenes from a bumpy boat or a surveillance camera tracking fast-moving vehicles on a highway often produces blurry images. This occurs because conventional RGB cameras (e.g., smartphones, surveillance systems, or instant cameras) capture images by integrating scene information over the exposure period.
	To address motion blur, various approaches have been proposed, typically modeling the blurry image as a latent clear image convolved with a blur kernel \cite{blurrymodel}. However, these methods often assume specific motion patterns, limiting their effectiveness in real-world scenarios where motion is complex and non-uniform.
	Event cameras offer a promising alternative for tackling motion blur. These bio-inspired sensors record pixel-level intensity changes asynchronously with high temporal resolution \cite{gallego2020event, xu2021motion, sun2022event}. By encoding intensity changes during the exposure time into an event stream, as shown in Fig. \ref{blurry_events}(c), event cameras provide a unique capability to assist in deblurring.
	
	Despite the impressive performance of event-based deblurring, event cameras, which respond to intensity changes with extremely high temporal resolution (on the microsecond scale), often generate a substantial volume of events during the exposure time of a blurry image. Each event, encoded using the Address Event Representation (AER) protocol, requires $8$ bytes \cite{AERprotocol}, comprising the triggered time $t$, pixel coordinates $(x,y)$, and polarity $p$. As a result, directly transmitting this large volume of events is highly bandwidth-costly. Furthermore, event-based deblurring methods rely on both blurry images and events as inputs, necessitating a transmission system capable of efficiently handling these two data types simultaneously. To address these challenges, we first consider the joint transmission of blurry images and events under a limited channel bandwidth, aiming to achieve high-quality reconstructions at the receiver.
	
	In this work, we propose a deblurring task-oriented semantic communication system that utilizes events as side information (EV-JSCC).
	To the best of our knowledge, this is the first semantic communication system designed specifically for deblurring tasks and the first to incorporate events into JSCC system design.
	Considering that RGB cameras and event cameras capture the same scene in distinct ways, they share overlapping information alongside their domain-specific details. Transmitting blurry images and events independently would result in redundant transmission of shared information. To avoid this, we design a shared encoder to extract and transmit the shared information, while employing an image encoder and an event encoder to capture their unique, domain-specific features. At the receiver, the received symbols are fed into a deblurring decoder for clear reconstructions. Specifically, our deblurring decoder is built upon the U-Net architecture \cite{unet}, with multiple cross-attention modules within the U-Net encoder to effectively fuse features from the blurry images and events. Additionally, to enhance the overall performance, we adopt a multi-stage training strategy. Simulation results demonstrate that our proposed system significantly outperforms existing JSCC methods in the joint transmission and deblurring task.
	
	\begin{figure*}[t]
		\begin{centering}
			\includegraphics[width=0.92 \textwidth]{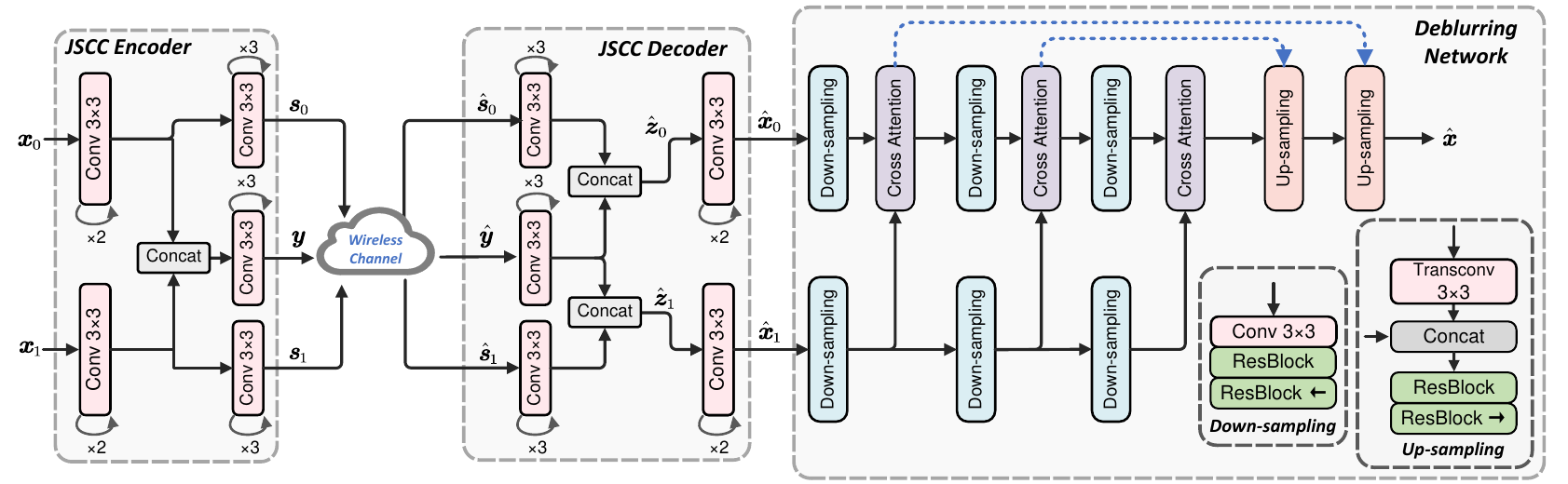}
			\par \end{centering}
		\caption{The architecture of the proposed EV-JSCC. }
		\label{archi}
	\end{figure*}
	
	\section{Framework of EV-JSCC}\label{SEC2}
	In this section, we present the framework of the proposed EV-JSCC system. To provide a comprehensive understanding, we begin by giving a brief introduction of blurry image, events and clear image, as well as their relations. 
	\subsection{Problem Analysis} \label{problem}
	As shown in Fig. \ref{blurry_events} (c), an event camera triggers events at pixel $(x,y)$ whenever the logarithm of the intensity changes beyond a pre-set threshold $c$, as described by the following equation:
	\begin{equation}\label{con:event_trigger}
		\log(\boldsymbol{I}_{xy}(t))-\log(\boldsymbol{I}_{xy}(t-\Delta t)) = p \cdot c,
	\end{equation}
	where $\boldsymbol{I}_{xy}(t)$ and $\boldsymbol{I}_{xy}(t-\Delta t)$ denote the instantaneous
	clear images at time $t$ and $t-\Delta t$ for a specific pixel location $(x, y)$, respectively. Here, $\Delta t$ is the time since the last event at this pixel location, and $p \in \left\lbrace +1,-1 \right\rbrace $ denotes the polarity, which indicates the direction of the intensity change (increase or decrease). Therefore, each event can be characterized by a tuple $(x,y, t, p)$. An example of events triggered during a period [$t_f-T/2$, $t_f+T/2$] is provided in Fig. \ref{blurry_events} (a). To facilitate the expression of events, we define $e_{xy}(t)$ as a continuous-time function for each pixel location $(x,y)$ in the image, such that
	\begin{equation}\label{event_with_dirac}
		e_{xy}(t) \triangleq  p \delta(t-t_0),
	\end{equation}
	whenever an event $ (x,y,t_0,p)$ occurs. Here we denote $\delta(\cdot)$ as the Dirac function. As a result, a sequence of discrete events can be transformed into a continuous-time signal.
	
	
	As shown in Fig. \ref{blurry_events}(a), the corresponding blurry image during the exposure interval [$t_f-T/2$, $t_f+T/2$] could be modeled as an average of clear latent intensity images $\boldsymbol{I}_{xy}$:		
	\begin{equation}\label{blur_model}
		\boldsymbol{B}_{xy}=\frac{1}{T} \int_{t_f-T/2}^{t_f+T/2} \boldsymbol{I}_{xy}(t) d t,
	\end{equation}
	where $t_f$ is the middle point of the exposure time $T$. We have $\log\left(\boldsymbol{I}_{xy}(t)\right)=\log\left(\boldsymbol{I}_{xy}(t_f)\right)+c \int_{t_f}^{t} e_{xy}(s) d s$ according to   \eqref{con:event_trigger} and   \eqref{event_with_dirac},
	then
	\begin{equation}\label{event_image_clear}
		\boldsymbol{B}_{xy} =\frac{\boldsymbol{I}_{xy}(t_f)}{T} \int_{t_f-T/2}^{t_f+T/2} \exp \left(c \int_{t_f}^{t} e_{xy}(s) d s\right) d t.
	\end{equation}
	As each pixel can be treated independently, the subscripts $x$ and $y$ are omitted hereafter. Finally, considering all pixels in an image, we derive a simple model that connects events, the observed blurry image, and the latent clear image:
	\begin{equation}\label{mathmodel}
		\boldsymbol{B} =\frac{\boldsymbol{I}(t_f)}{T} \int_{t_f-T/2}^{t_f+T/2} \exp \left(c \int_{t_f}^{t} e(s) d s\right) d t,
	\end{equation}
	this formulation suggests that the latent clear image $\boldsymbol{I}(t_f)$ can be derived from the blurry image combined with the set of events $\mathcal{E} = \{e_i = (x_i, y_i, t_i, p_i): t_f - T/2 \leq t_i \leq t_f + T/2\}$, which includes all events triggered during the exposure time. By accumulating these events over time, it becomes possible to derive the clear image. To enable this reconstruction at the receiver, we propose the EV-JSCC method for the joint transmission of blurry images and events, as detailed in Section II-B.
	
	\subsection{System Model} \label{system}
	In this work, we focus on transmission tasks in a hybrid camera system, comprising a time-synchronized and spatially-synchronized RGB camera and event camera, ensuring the same scene is captured within the same period. The outputs generated by the hybrid camera system—the blurry image $\bm{x}_0 \in \mathbb{R}^{n_0}$, where $n_0$ is the dimension of the input image, as well as the event stream, $\bm{x}_1 \in \mathbb{R}^{n_1}$, where $n_1$ represents the size of the event representation (further detailed in Section \ref{SEC3})—serve as the two inputs for our proposed system. Given the heavy overhead for transmitting events, we conceive of a deblurring task-oriented JSCC system, which consists of an image encoder, an event encoder, a shared encoder, an image decoder, an event decoder, and a deblurring decoder, as illustrated in Fig. \ref{framework}.
	
	The image encoder $f_{\bm{\theta}_0}: \mathbb{R}^{n_0}\rightarrow \mathbb{C}^{k_0}$ maps $\bm{x}_0$  to a complex vector $\bm{s}_0 \in \mathbb{C}^{k_0}$, where $\bm{\theta}_0$ denotes the parameters of the image encoder and $k_0$ denotes the number of symbols to be transmitted. 
	The event encoder $f_{\bm{\theta}_1}: \mathbb{R}^{n_1}\rightarrow \mathbb{C}^{k_1}$ maps $\bm{x}_1$  to a complex vector $\bm{s}_1 \in \mathbb{C}^{k_1}$, where $\bm{\theta}_1$ denotes the parameters of the event encoder and $k_1$ denotes the output dimension.
	The shared encoder $f_{\bm{\theta}_y}: \mathbb{R}^{n_0+n_1}\rightarrow \mathbb{C}^{k_2}$, fuses the features of $\bm{x}_0$ and $\bm{x}_1$, and maps the fused features to a complex vector $\bm{y} \in \mathbb{C}^{k_2}$, where $\bm{\theta}_y$ denotes the paremeters of the shared encoder and $k_2$ denotes the number of transmitted symbols.
	Accordingly, the channel bandwidth ratio (CBR) is defined as $\rho = (k_0+k_1+k_2)/n_0$, denoting the channel uses per image pixel. The encoding process is given by:
	\begin{equation}
		\bm{s}_0 = f_{\bm{\theta}_0}(\bm{x}_0),
		\bm{s}_1 = f_{\bm{\theta}_1}(\bm{x}_1),
		\bm{y} = f_{\bm{\theta}_y}(\bm{x}_0, \bm{x}_1).
	\end{equation}
	
	Before transmission, we concatenate all encoders' outputs to obtain the transmitted symbol stream $\bm{z} = [\bm{s}_0, \bm{s}_1, \bm{y}]$. To account for the limited energy in real-world communication systems, we impose an average power constraint on $\bm{z}$. The channel input symbol stream $\bm{z}$ is then transmitted through a noisy wireless channel, denoted by $\eta: \mathbb{C}^{k}\rightarrow\mathbb{C}^{k}$, where $k=k_0+k_1+k_2$. In this work, an AWGN (additive white Gaussian noise) channel is adopted, thus the transmission process is modeled as follows:
	\begin{equation}
		\bm{\hat{z}} = \bm{z} + \bm{n},
	\end{equation}
	where $\bm{n} \sim \mathcal{CN}(0,\sigma^2 \bm{I}_{k \times k})$ is a complex Gaussian vector with variance $\sigma^2$. 
	
	At the receiver, $\bm{\hat{z}}$ is first separated to $\bm{\hat{s}}_0, \bm{\hat{s}}_1$ and $\bm{\hat{y}}$. Then, $\bm{\hat{y}}$ is concatenated with $\bm{\hat{s}}_0$ and $ \bm{\hat{s}}_1$ to produce $\bm{z}_0$ and $\bm{z}_1$, which correspond to the features of the blurry image and the events, respectively. These features are subsequently decoded by the image decoder $g_{\bm{\phi}_0}:\mathbb{C}^{k_0+k_2} \rightarrow \mathbb{R}^{n_0}$ and the event decoder $g_{\bm{\phi}_1}:\mathbb{C}^{k_1+k_2} \rightarrow \mathbb{R}^{n_1}$, yielding the reconstructed image $\bm{\hat{x}}_0$ and reconstructed events $\bm{\hat{x}}_1$. The process is given by:
	\begin{equation}
		\bm{\hat{x}}_0 = g_{\bm{\phi}_0}(\bm{\hat{z}}_0) \in \mathbb{R}^{n_0},
		\bm{\hat{x}}_1 = g_{\bm{\phi}_1}(\bm{\hat{z}}_1) \in \mathbb{R}^{n_1}.
	\end{equation}	
	Finally, the reconstructed image and events are processed through a deblurring decoder $g_{\bm{\phi}_y}:\mathbb{R}^{n_0+n_1} \rightarrow \mathbb{R}^{n_0}$ to recover the clear image:
	\begin{equation}
		\bm{\hat{x}} = g_{\bm{\phi}_y}(\bm{\hat{x}}_0, \bm{\hat{x}}_1) \in \mathbb{R}^{n_0},
	\end{equation}	
	where $\bm{\hat{x}}$ denotes the final clear image.
	%
	
	\section{Model Architecture and Training strategy}\label{SEC3}
	
	In this section, we detail the architecture and the multi-stage training strategy of the proposed EV-JSCC.
	
	\subsection{Unified Representation of Events}
	We design a unified representation of events for two main reasons. Firstly, the large number of events triggered during the exposure time of a blurry image makes transmission highly costly. Additionally, the number of events may vary across different images, posing a challenge for neural networks, which typically require fixed-dimensional inputs. To address this, a unified representation is essential.
	
	Inspired by \cite{sun2022event}, we divide the exposure time $T$ of the blurry image into $2K$ intervals, resulting in $2K+1$ boundaries. Ler $\boldsymbol{S}_j(x,y)$ represent the integral of the events from the middle of exposure time $t_f$ to the $j$-th boundary, where $k=0,1,...,2K$:
	\begin{equation}
		\boldsymbol{S}_j(x,y) = \int_{t_f}^{t_f+ (k-K)\frac{T}{2K}} e_{xy} d t,
	\end{equation}	
	where $j \textgreater K$ corresponds to the difference between the positive and negative events, and $j \textless K$ corresponds to the difference between the negative and positive events. This process transforms the event stream into a tensor with the shape $2K \times H \times W$, where $H \times W$ corresponds to the resolution of the event camera. We set $K=3$ in our experiments.
	
	\subsection{Shared and Domain-Specific Features Extraction}
	As shown in Fig. \ref{archi}, we use $2$ CNN layers to transform each high-dimensional input into a low-dimensional feature. Considering that both RGB cameras and event-based cameras capture the same scenario in distinct manners, they share some common information while retaining domain-specific characteristics. Transmitting these inputs independently could lead to inefficient use of channel bandwidth. To address this issue, we design a shared encoder to extract the shared information between the blurry image and events, an image encoder, and an event encoder to capture their domain-specific features. Each encoder consists of $3$ CNN layers.
	At the receiver, we first employ $3$ CNN layers for each received signal. Afterward, we combine the shared signal with the domain-specific signal, producing image features and event features, respectively. These features are then passed through $2$ CNN layers to reconstruct the blurry image and events, respectively. 
	
	\subsection{Deblurring Decoder} \label{cluster}
	As modeled in Section \ref{problem}, the latent clear image can be recovered from the blurry image and its corresponding events. To achieve this, we build a deblurring decoder that integrates both as inputs. This module, shown in Fig. \ref{archi}, is built upon the U-Net architecture \cite{unet}. 
	The deblurring decoder begins by extracting hierarchical features from the blurry image and events using a U-Net encoder, which consists of multiple down-sampling units. During this process, the features from the blurry image are fused with the corresponding event features at each level before being passed to the next feature extractor. To enhance this fusion, we incorporate a cross-attention block \cite{sun2022event} at each level, enabling effective interaction between the blurry image and events.
	The U-Net decoder, composed of several up-sampling units, progressively refines the details of the reconstructed image. Skip connections (illustrated by blue dotted lines) link the outputs of the down-sampling units directly to their corresponding up-sampling units, preserving critical spatial information throughout the decoding process. This architecture effectively leverages the complementary information from both the blurry image and events, enabling the network to generate a clear image with high fidelity.
	\begin{algorithm}[t] 
		\caption{Training the EV-JSCC} 
		\label{Training}
		\SetAlgoLined
		\textbf{Input:} A hybrid dataset $\mathcal{X}$ of blurry images along with corresponding events, and the learning rate $l_r$. \\
		\textbf{Output:} Parameters $\left(\bm{\theta}_0^*, \bm{\theta}_1^*, \bm{\theta}_y^*, \bm{\phi}_0^*, \bm{\phi}_1^*, \bm{\phi}_y^* \right)$. \\
		\textbf{First stage: Train transmission modules.}\\
		Randomly initialize the parameters $(\bm{\theta}_0, \bm{\theta}_1, \bm{\theta}_y, \bm{\phi}_0, \bm{\phi}_1)$ and freeze the parameters $\bm{\phi}_y$. \\
		\For{each epoch}{
			Sample $(\bm{x}_0, \bm{x}_1)$ from hybrid dataset $\mathcal{X}$. \\
			Calculate the loss function based on (\ref{loss_func_0}). \\
			Update the parameters $(\bm{\theta}_0, \bm{\theta}_1, \bm{\theta}_y, \bm{\phi}_0, \bm{\phi}_1)$.\\
		} 
		------------------------------------------------------------ \\
		\textbf{Second stage: Train deblurring decoder.}\\
		Load and freeze the parameters $(\bm{\theta}_0, \bm{\theta}_1, \bm{\theta}_y, \bm{\phi}_0, \bm{\phi}_1)$ from the first stage and initialize the parameters $\bm{\phi}_y$. \\
		\For{each epoch}{
			Sample $(\bm{x}_0, \bm{x}_1)$ from hybrid dataset $\mathcal{X}$. \\
			Calculate the loss function based on (\ref{loss_func_1}). \\
			Update the parameters $\bm{\phi}_y$.\\
		}
		------------------------------------------------------------ \\
		\textbf{Train the whole model.}\\
		Load the parameters $(\bm{\theta}_0, \bm{\theta}_1, \bm{\theta}_y, \bm{\phi}_0, \bm{\phi}_1)$ from the first stage and the parameters $\bm{\phi}_y$ from the second stage. \\
		\For{each epoch}{
			Sample $(\bm{x}_0, \bm{x}_1)$ from hybrid dataset $\mathcal{X}$. \\
			Calculate the loss function based on (\ref{loss_func_1}). \\
			Update the parameters $(\bm{\theta}_0, \bm{\theta}_1, \bm{\theta}_y, \bm{\phi}_0, \bm{\phi}_1, \bm{\phi}_y)$.\\
		}
	\end{algorithm}

	\begin{figure*}[t]
		\begin{centering}
			\includegraphics[width=0.78 \textwidth]{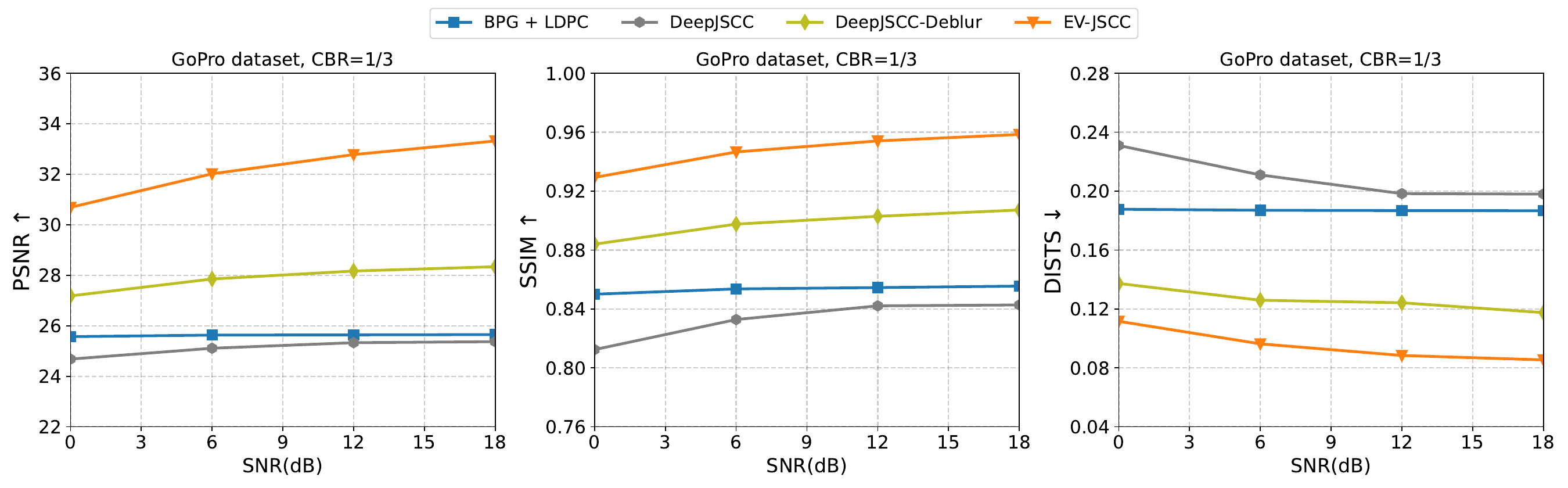}
			\par \end{centering}
		\caption{The performance of the proposed model on the GoPro dataset at a CBR of $1/3$ versus SNR.}
		\label{gopro}
	\end{figure*}
	
	\begin{figure*}[t]
		\begin{centering}
			\includegraphics[width=0.78 \textwidth]{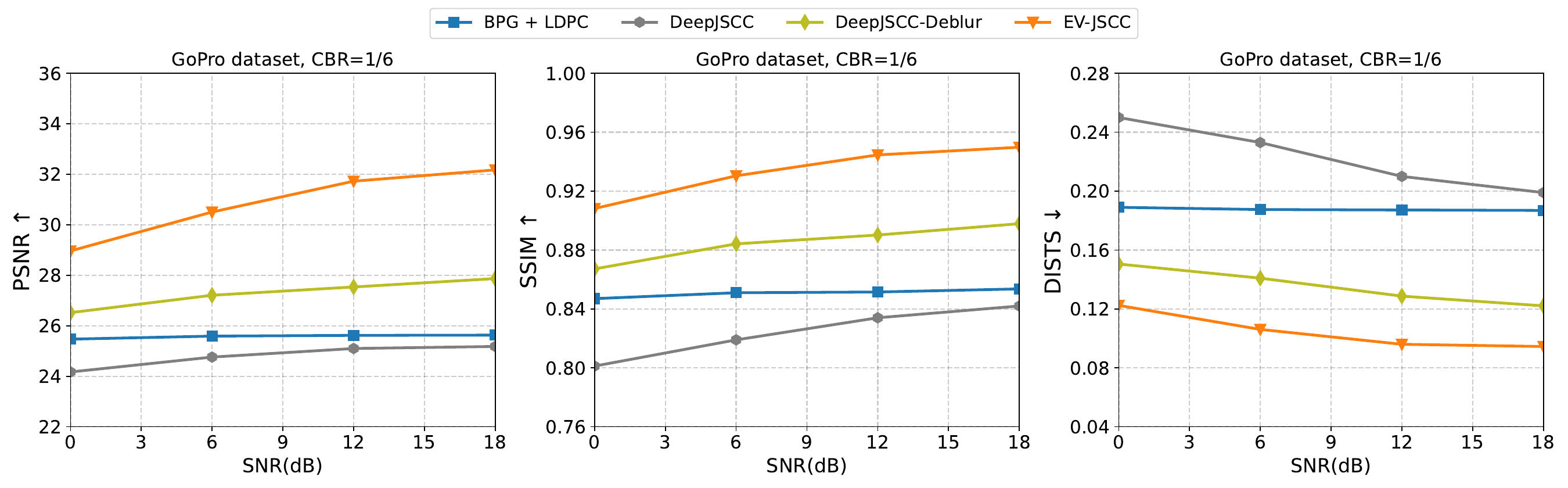}
			\par \end{centering}
		\caption{The performance of the proposed model on the GoPro dataset at a CBR of $1/6$ versus SNR.}
		\label{reblur}
	\end{figure*}
	\subsection{Training Strategy} \label{train}
	To ensure training stability and enhance overall performance, we propose a multi-stage training strategy: 
	\begin{itemize}
		\item Train transmission modules: we initially freeze the parameters of the deblurring decoder and train other modules, with the loss function defined as:
		\begin{equation} \label{loss_func_0}
			\begin{aligned}
				\mathcal{L}=& \mathbb{E}_{\bm{x}_0, \bm{x}_1}[||\bm{x}_0-{\bm{\hat{x}}}_0||^2 + ||\bm{x}_1-\bm{\hat{x}}_1||^2],
			\end{aligned}
		\end{equation}
		where $\mathbb{E}_{\bm{x}_0, \bm{x}_1}||\bm{x}_0-\bm{\hat{x}}_0||^2$ is the mean-square error (MSE) between blurry image $\bm{x}_0$ and its reconstruction $\bm{\hat{x}}_0$, and  $\mathbb{E}_{\bm{x}_0, \bm{x}_1}||\bm{x}_1-\bm{\hat{x}}_1||^2$ represents the MSE between events $\bm{x}_1$ and recovered events $\bm{\hat{x}}_1$.
		\item Train deblurring decoder: we freeze the parameters trained in the first stage and train the deblurring decoder individually, employing the loss function defined as:
		\begin{equation} \label{loss_func_1}
			\begin{aligned}
				\mathcal{L}=& \mathbb{E}_{\bm{x}_0, \bm{x}_1}||\bm{x}-\bm{\hat{x}}||^2,
			\end{aligned}
		\end{equation}
		where $\mathbb{E}_{\bm{x}_0, \bm{x}_1}||\bm{x}-{\bm{\hat{x}}}||^2$ is the MSE between the ground-truth clear image $\bm{x}$ and the deblurred image $\bm{\hat{x}}$.
		\item Train the whole model: we fine-tune the entire model with the loss function defined in Eq. (\ref{loss_func_1}).
	\end{itemize}
	
	The multi-stage training process is outlined in Algorithm \ref{Training}.	
	\section{Simulation Results} \label{SEC4}
	In this section, we perform simulations to evaluate the performance of our proposed models.
	\subsection{Simulation Settings}
	\subsubsection{Basic Settings}
	All experiments are conducted using Pytorch. The Adam optimizer is utilized to perform stochastic gradient descent. We initially set the learning rate to $2\times 10^{-4}$ with plans to reduce it after several epochs. The training process is conducted over $800$ epochs with a batch size of $8$.
	
	\subsubsection{Datasets}
	We evaluate our scheme on the GoPro dataset \cite{gopro}, which consists of $3,214$ pairs of blurry images and clear images, with a resolution of $1,280\times720$. Specifically,  the blurry images are generated by averaging multiple high-speed clear images. To simulate events, we resort to ESIM \cite{esim}, an open-source event camera simulator, to generate corresponding events for images.
	
	
	\subsubsection{Benchmarks}
	We have considered three methods as benchmarks: 
	\begin{itemize}
		\item DeepJSCC: The network is essentially the same as the DeepJSCC \cite{Eirina_TCCN2019}, where it transmits blurry images at the transmitter, with the loss function defined as the MSE between the reconstructed image and the ground-truth clear image.
		\item DeepJSCC-Deblur:  This method adopts the same settings as the benchmark DeepJSCC, with the exception that the deblurring module is introduced at the receiver. Notably, the event input to the deblurring module is set to a constant tensor, meaning no events are introduced during the process.
		
		\item BPG+LDPC: This method employs better portable graphics (BPG) for source coding and  low-density parity-check (LDPC) for channel coding, followed by
		quadrature amplitude modulation (QAM).
	\end{itemize} 
	\begin{figure*}[t]
		\begin{centering}
			\includegraphics[width=0.89 \textwidth]{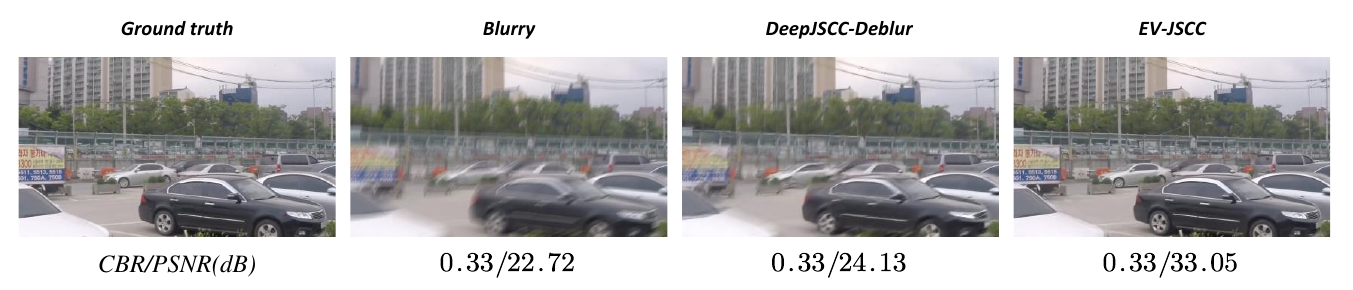}
			\par \end{centering}
		\caption{Visualization example of the reconstructed images, where the metrics are [CBR/PSNR]. These reconstructed images are obtained by employing different schemes over the AWGN channels at the SNR of 18 dB.}
		\label{visualization}
	\end{figure*}			
	\subsubsection{Considered Metrics}
	We evaluate the performance of the methods using $3$ metrics, including peak signal-to-noise ratio (PSNR), structure similarity index measure (SSIM), and deep image structure and texture similarity (DISTS).
	PSNR is a widely used metric for assessing image quality by calculating pixel-wise distortion. SSIM measures the similarity between two images by analyzing luminance, contrast, and structural features. DISTS employs a pre-trained network to extract high-dimensional features from two images, and the distance between them is defined as DISTS.
	
	\subsection{Performance Comparision}

	We evaluate the performance of our proposed EV-JSCC on the AWGN channels. In all subsequent experiments, the
	model is trained in a single SNR regime and tested in the same SNR regime.
	
	Fig. \ref{gopro} compares our EV-JSCC with benchmarks on the GoPro dataset at a CBR of $1/3$. The results demonstrate significant advantages of our proposed EV-JSCC. Clearly, our method and DeepJSCC-Deblur outperform DeepJSCC and BPG+LDPC. This enhancement is primarily attributed to the powerful deblurring module. This highlights the advantages of semantic communications, where task-oriented features are transmitted to optimize channel bandwidth usage, resulting in significant performance enhancements for the specific task.
	We also noticed that the performance of the BPG+LDPC method is almost unaffected by varying channel conditions. This is because the transmitted blurry image has already lost many details compared to the ground-truth clear image, so even if errors or damage occur during transmission, the degradation in image quality remains minimal and less noticeable.
	Notably, our proposed method exhibits significant advantages over DeepJSCC-Deblur across all SNR levels. This improvement is primarily attributed to the incorporation of events and the specialized JSCC encoder and decoder, which effectively extract both shared and domain-specific information.
	
	Fig. \ref{reblur} compares the performance of our proposed EV-JSCC with benchmark methods at a CBR of $1/6$. The results align with those observed at a CBR of $1/3$, as shown in Fig. \ref{gopro}, further emphasizing the substantial superiority of EV-JSCC and its generalization.
	
	To provide a more intuitive understanding of the performance, Fig. \ref{visualization} showcases examples of deblurred images produced using various transmission schemes. Compared to the best benchmark DeepJSCC-Deblur, our EV-JSCC delivers significantly better visual quality, preserving more texture and fine details. This further demonstrates the potential of employing
	the proposed methods in wireless image transmission and deblurring.
	
	
	\section{Conclusion} \label{SEC5}
	In this work, we proposed a deblurring task-oriented semantic communication system that utilizes events as side information. 
	To avoid transmitting redundant information, we designed a shared encoder to extract and transmit the shared features between the blurry image and events, complemented by an image encoder and an event encoder to capture domain-specific features. 
	At the receiver, the transmitted symbols are processed through a deblurring decoder based on the U-Net architecture, enabling high-quality reconstructions. 
	Additionally, we implemented a multi-stage training strategy to enhance the overall performance. Simulation results demonstrated that our proposed system significantly outperforms existing JSCC methods in joint transmission and deblurring tasks.
	
	\bibliographystyle{IEEEtran}
	\bibliography{IEEEabrv,Reference}
	
\end{document}